\documentclass{iopart}
\usepackage{graphicx}
\usepackage{cite}
\usepackage{iopams}

\begin{document}

\title[Three delta-function wells giving rise to EP3s]
{A model of three coupled wave guides and third order exceptional points}

\author{  W D Heiss$^{1,2}$, G Wunner$^{3}$,}

\address{$^1$Department of Physics, University of Stellenbosch,
  7602 Matieland, South Africa}
\address{$^2$National Institute for Theoretical Physics (NITheP), Western Cape,\
  South Africa}
\address{$^3$Institut f\"ur Theoretische Physik, Universit\"at Stuttgart,
  Pfaffenwaldring 57, 70\,569 Stuttgart, Germany}
\ead{dieter@physics.sun.za}

\ead{wunner@itp1.uni-stuttgart.de}
\begin{abstract}
A ${\cal PT }$-symmetric model for three interacting wave guides is investigated. Each wave guide is represented by an
attractive $\delta$-function potential being in equidistant positions. The two outer potentials are complex
describing loss and gain, respectively. The real parts of the outer potentials are assumed to be equal.
The major focus of the study lies on the occurrence of an exceptional point of third order and
the physical effects of such singularity. While some results resemble those from similar studies
with two wave guides, the three wave guides appear to have a richer structure. Emphasis is placed
on the fine tuning in the approach of the EP3 as this appears to be a particular challenge for
an experimental realization.

\end{abstract}
 
\submitto{\JPA}
\maketitle

\section{\label{sec:intro} Introduction}


Exceptional points are points in the parameter space of 
a physical system where both eigenvalues and eigenvectors coincide.
They can occur only for non-hermitian operators.
Exceptional points appear in particular in $\cal PT$-symmetric systems
which are symmetric under the combined action
of parity inversion and time reversal. In a ground-breaking paper in 1998
Bender and Boettcher \cite{Bender98} demonstrated that $\cal PT$-symmetric 
non-hermitian Hamiltonians can have real eigenvalues, and that eigenstates coalesce at exceptional
points when the symmetry is broken. Since then there has been a host of papers discussing $\cal PT$ symmetry
in a diversity of physical systems, involving microwave cavities, superconductivity,
atomic diffusion, nuclear magnetic resonance, coupled classical and electronic oscillators, and in
particular in optics (see, e.\,g., \cite{Moiseyev2011a, Heiss12, Bender13} and references therein).

In a seminal paper in 2008, Klaiman et al. \cite{klaiman08a} proposed the experimental
visualization of exceptional points of second order in a system of two coupled 
$\cal PT$-symmetric wave guides. Their model consists of two wave guides in which the refractive index
differs from that of the background substrate ($n_0$) by a small amount $\Delta n$, and imaginary 
parts $\pm \gamma $  of equal size but opposite sign are introduced in the guides. Their 
predictions received convincing confirmation 2010 in the experiment by R\"uter et al.
\cite{rueter10}, when the coalescence of two wave guide modes  at a branch point of 
second order was observed, when gain and loss were increased up to a critical value. 
In the experiment, loss is realized by pasting a metal on one wave guide, and gain by
pumping laser light on the other.

The present paper goes beyond these investigations and explores the 
possibility of observing exceptional points of higher order, i.e. the coalescence of more
than two modes in multi wave guide systems. Specifically, as a natural next step we consider a 
$\cal PT$ triple wave guide system and search for the physical conditions under which an 
exceptional point of third order (EP3) could be observed. The effects of exceptional 
points of higher order in particular of third order (EP3s),  have received increasing attention
in recent years 
\cite{Graefe08a, Heiss08, Graefe12a, Ryu12, Gutoehrlein13, Heiss2015,Am-Shallem15,
Am-Shallem15b,Ding15}. Extending the model by Klaiman et
al. \cite{klaiman08a} we place a third wave guide between the guides with gain  and 
loss, but with only a real part of the refractive index that may be different
from that of the outer wave guides.

We make use of the formal analogy between the equation of electrodynamics 
in paraxial approximation governing the propagation of waves and the one-dimensional
Schr\"odinger equation of quantum mechanics. In this analogy the propagation direction
of the waves (usually the $z$ direction) is supplanted by time, and the refractive index
$n(x)$ is replaced by the potential $V(x) = - k_0^2 n^2(x)$, where $k_0$ is the vacuum wave number.
Thus the equivalent quantum mechanical problem is that of two potential wells of equal depth and the 
same amount of gain in one and loss in the other, and a third well with only a real-valued 
potential of different depth between them.

To gain insight we simplify the problem further and model the potential wells by three
delta functions. Delta-function potentials are popular  as model 
systems in the literature \cite{Jakubsky05, Mostafazadeh2006a, Mehri-Dehnavi2010a, 
Mostafazadeh13b, Jones2008a, Cartarius12b, Cartarius12c, Loehle15, Barashenkov2015},
since they allow for analytic or partially analytic solutions, 
but are flexible enough to provide insight into characteristic phenomena of the more
complex physical situations. Our model is expected to capture the essential features of the 
real, i.e.~experimental problem, and may serve as a guide for the search of higher exceptional 
points in real multi wave guide systems. In fact, our findings point to high sensitivity in
the parameters near to the EP3. In particular it is argued that, in contrast to the case of an EP2,
a close approach of an EP3 cannot be achieved with only one real parameter. We hope that our findings
can serve as a guide in an experimental effort to show that an EP3 is a physical reality.

\section{The model for three wave guides}

We model a $\cal PT$-symmetric system of three coupled wave guides 
by three delta-function
shaped potential wells located at $x = \pm b$ and $x=0$, where loss is added to the left
well while the same amount of gain is added to the right well. The connection to realistic
wave guides is established in \ref{waveguides}.
The corresponding Schr\"odinger equation used in the present paper reads: 

\begin{eqnarray}\label{SE}
-\Psi^{\prime\prime}(x) &-&\left[(1+i\gamma)\delta(x+b) + 
\Gamma \delta(x) +
(1-i\gamma)\delta(x-b)\right]\Psi(x) \nonumber \\
&=& -k^2 \Psi(x)\, .
\end{eqnarray}

The real-valued parameter
$\gamma$ determines the strength of the gain and loss terms.
Units have been chosen in such a way that the strength
of the real part  of the two outer delta-function potentials  is normalized to unity,
while in the middle well we allow for a different depth given by the the real
parameter $\Gamma > 0$.
For stationary solutions the eigenvalues $k$ are real,
but since the complete eigenvalue spectrum is complex in general,
we will also consider solutions 
with $k \in {\mathbb {C}}, \; {\rm Re}(k) > 0$. Yet our major emphasis is focused
upon the bound state solutions with real eigenvalues.

The bound-state wave function has the form:
\[
\Psi(x) = \left\{ \begin{array}{l@{\quad:\quad}l}
A\, {\rm e}^{k x} & x < -b \\[0.4ex]
 2\,\left(r \cosh (kx) + \varrho_1 \sinh (kx)\right) & -b < x < 0 \\[0.4ex]
2\, \left(r \cosh (kx) + \varrho_2 \sinh (kx)\right) & 0 < x < b  \\[0.4ex] 
B\, {\rm e}^{- k x} & b < x
\end{array} \right.
\]

Applying at the delta functions the continuity conditions for the wave functions and the discontinuity conditions for their 
first derivatives we obtain the system of linear equations 
\begin{equation}
{\cal M} 
\begin{pmatrix} 
{r \cr \varrho_1 \cr \varrho_2
}
\end{pmatrix}
=
\begin{pmatrix} 
{0 \cr 0 \cr 0
}
\end{pmatrix}
\label{lineq}
\end{equation}
with the matrix 

\begin{eqnarray} \qquad \qquad \qquad {\cal M } = \nonumber   \\
\nonumber  \\
\begin{pmatrix} 
 {\kappa_0 {\rm e}^{-2kb}  + \kappa_0 -2k & \kappa_0 {\rm e}^{-2kb} -\kappa_0 + 2k & 0 \cr 
\kappa_0^\ast {\rm e}^{-2 k b} + \kappa_0^\ast - 2 k & 0 & - \kappa_0^\ast {\rm e}^{-2 k b} + \kappa_0^\ast - 2 k \cr
-\Gamma & k & -k
}
\end{pmatrix} 
\label{matrix}
\end{eqnarray}
where $\kappa_0 = 1 + i \gamma$.
The remaining wave function coefficients are related to $r, \varrho_1, \varrho_2$ by $A = r(1 + {\rm e}^{2kb}) + \varrho_1 (1 - {\rm e}^{2kb})$, $B = r(1 + {\rm e}^{2kb}) + \varrho_2 ({\rm e}^{2kb} - 1)$.

The eigenvalues $k$ are obtained by finding the roots of 
the corresponding secular equation
\begin{eqnarray}
\qquad \qquad \det({\cal M})&=&  \nonumber \\
  \Gamma \Big( {\rm e}^{-4 k b} (1+ \gamma^2 ) &-& 2 {\rm e}^{-2 k b} (\gamma^2- 2 k +1) + \gamma^2 + (2 k-1)^2 \Big)  +    
\nonumber \\
+ 2 k \Big( {\rm e}^{-4 k b} (1+ \gamma^2 )  &-& \gamma^2 - (2 k-1)^2 \Big) = 0 \;.
\label{sec}
\end{eqnarray}
The roots depend on three parameters, the distance $b$ of the wells, the
strength $\gamma$ of the loss/gain terms, and the depth $\Gamma$ of the middle well.

It is instructive to consider the
limit  $k b \gg 1$, i.~e., no coupling between the modes. Then (\ref{sec}) simplifies
to 
\begin{equation}
\left[
 \gamma^2 + (2 k - 1 )^2
\right)] 
( \Gamma - 2k ) = 0
\end{equation}
 with the solutions $k_1 = \Gamma/2$ and $k_{2,3} = (1 \pm i \gamma)/2$. This means that the
 middle well retains its unperturbed eigenvalue, while the eigenvalues of the outer
 wells acquire an imaginary part, corresponding to the exponential growth and decrease of the 
 gain and loss mode, respectively. This demonstrates that $\cal PT$ symmetric modes
 can exist only when there is sufficient coupling between the wave guides. 
 
Within the present context the case $\gamma =0$ requires special treatment for the eigenvector of
the intermediate (in size), i.e.~second eigenvalue $k_2$. In fact, the
explicit values for $\varrho_{1,2}^{(2)}$ obtained for $\gamma >0$ blow up when $\gamma \to 0$ (see \ref{coeff}).
It is related to the fact that in the limit $\gamma \to 0$, the determinant of ${\cal M}$ (see (\ref{sec})) factorizes as
\begin{equation}
({\rm e}^{-2kb} +2k-1)\left( \Gamma({\rm e}^{-2kb} +2k-1) + 2k ({\rm e}^{-2kb}- 2k+1)\right) = 0 \;,
\label{gamma0}
\end{equation}
which for the second root implies ${\rm e}^{-2k_2b} +2k_2-1 = 0$. Note that this second eigenvalue  -- yielding the relation $b=-{\rm Log}(1-2k_2)/(2k_2)$ --
is independent of $\Gamma $. Inserting this expression for $b$ into that for $\varrho_{1,2}^{(2)}$ we obtain the expansion
\begin{equation}
 \varrho_{1,2}^{(2)} =-i \frac{1-2k_2}{\gamma \,k_2} \pm \frac{1-k_2}{k_2}+O(\gamma )^2
\end{equation}
and hence
\begin{equation}
\lim_{\gamma \to 0} \gamma \; \varrho _{1,2}^{(2)}= -i\frac{1-2k_2}{k_2}
\end{equation}
from which follows
\begin{equation}
 \lim_{\gamma \to 0} \gamma \begin{pmatrix}{r \cr \varrho_1^{(2)} \cr \varrho_2^{(2)} } \end{pmatrix}  = 
-i\frac{1-2k_2}{k_2}\begin{pmatrix}{0 \cr 1 \cr 1} \end{pmatrix}.
\end{equation}
The essential finding is the factor $-i$ for this second eigenvector 
in the limit $\gamma \to 0$. As we see in the following section it is due to this factor that we obtain a smooth
dependence on $\gamma $ also for the eigenvector associated with the second eigenvalue.

\section{Results and Discussion}
\subsection{Eigenvalues}
For $\gamma =0$ the model is hermitian, i.e.~we expect three real eigenvalues. 
Owing to the underlying ${\cal PT}$ symmetry we expect real eigenvalues for some range $\gamma >0$ until we reach
a coalescence of at least two eigenvalues where two eigenvalues become complex at an EP2. Such value depends on the
other parameters $b$ and $\Gamma $. Here we seek these parameters in such a way that the triplet of the three (real) parameters
($\gamma,\;b,\;\Gamma$) leads to the coalescence of all three levels, i.e.~to an EP3. \\
Let us recall that if a non-hermitian
Hamiltonian has an EP3 at a real eigenvalue a change of only one parameter would result in the sprouting out of three
coalescing eigenvalues in a symmetric way meaning that at most one eigenvalue can be real while the other two will be
complex. In other words, to achieve the coalescence of three real eigenvalues at least two parameters have to be
judiciously chosen. It is at this point where we would expect the need for a careful fine-tuning in an experimental realization.

\begin{figure}[t]
\begin{center}
\vspace*{-3mm}
\includegraphics[width=0.4\columnwidth]{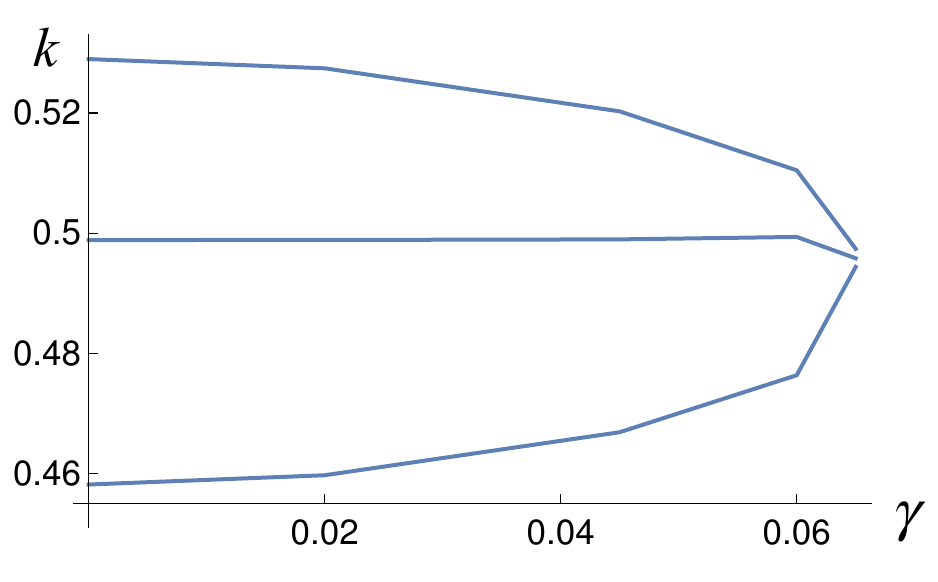}
\end{center}
\caption{Eigenvalues $k$ for a few points of $\gamma $ of the three bound states of the
underlying model for $b = 6.1$.
Note that close to the exceptional point where the three eigenvalues would
coalesce the distance has to be adjusted to $b = 6.2075$ (see text).}
\label{fig_1}
\end{figure}

In Fig.1 a set of three eigenvalues is illustrated as a function of $\gamma $ for $b=6.1$ and $\Gamma =1.002$.
Note that the second eigenvalue is almost unaffected by the variation of $\gamma $ while the other two eigenvalues
show a rather weak dependence except for the immediate neighbourhood of the EP3; the illustration stops short of the EP3.
The weak dependence on $\gamma $ of the eigenvalues -- except within the near neighbourhood of the EP3 -- is due to the
fairly large value of $b$: the states interact weakly away from the singularity.
For the whole interval $0\le \gamma \le 0.06$ the distance between the $\delta$-functions is $b=6.1$.
In line with the previous paragraph we had to re-adjust to $b=6.2075$ for $\gamma =0.065$ to ensure
real eigenvalues. Had we chosen this value for $b$ also
for $\gamma \le 0.06$ an EP2, i.e.~a coalescence of two pairs or complex eigenvalues would have resulted. The values at the EP3 are
$\gamma = 0.065278,\;b = 6.20124,\;k = 0.495849$ for $\Gamma=1.002$.

\begin{figure}[t]
\begin{center}
\vspace*{-3mm}
\includegraphics[width=0.8\columnwidth]{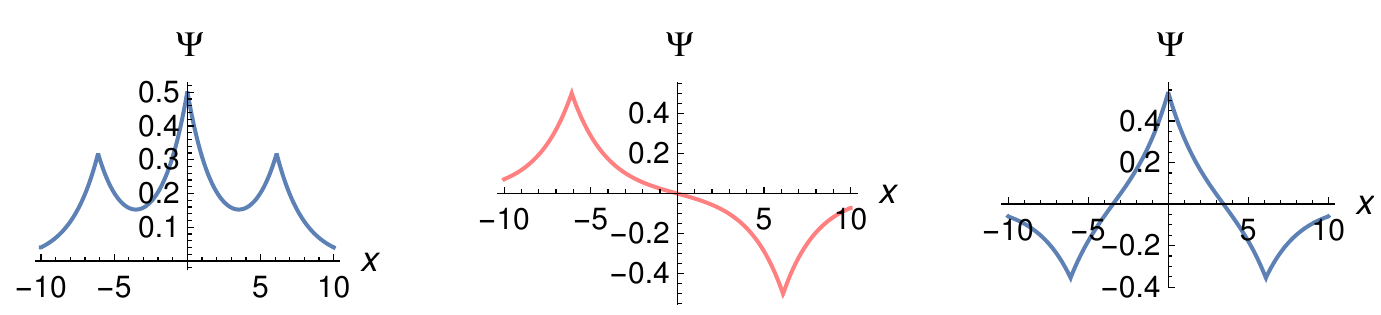}
\includegraphics[width=0.8\columnwidth]{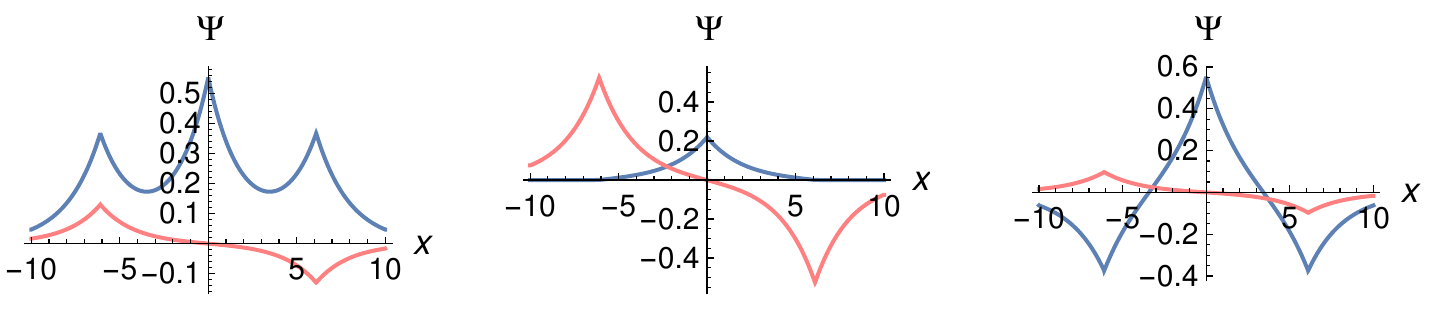}
\includegraphics[width=0.8\columnwidth]{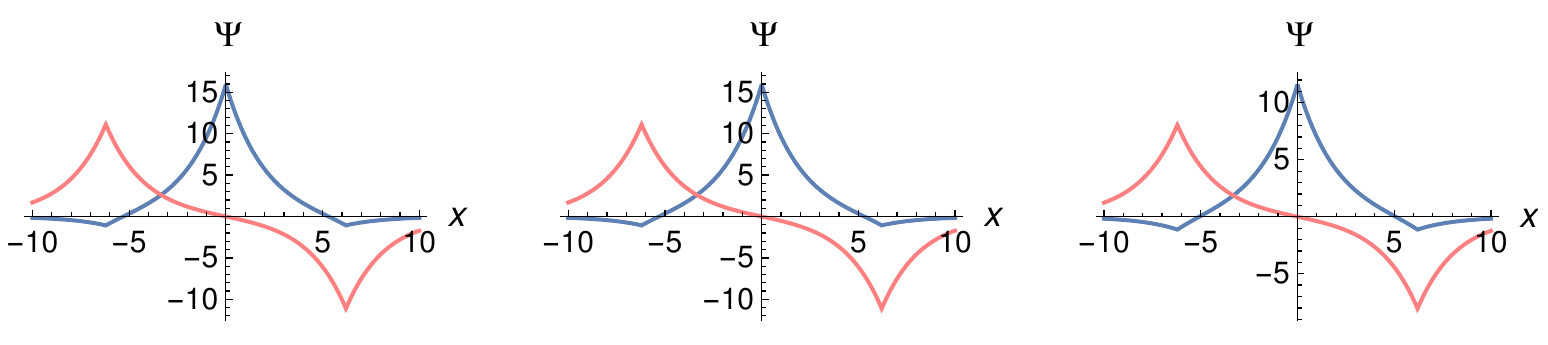}
\end{center}
\caption{C-normalized wave functions (see text) of the bound states of the underlying model
for increasing non-hermiticity parameter (from top to bottom) $\gamma = 0,\; 0.02$
and $0.065$, blue and red illustrate real and imaginary parts, respectively.
The rows correspond to decreasing eigenvalues from left to right.
The distance between the wells in the two upper panels is $b = 6.1$, 
while in the bottom panel, close to the EP3, it is $b = 6.2075$.}
\label{fig_2}
\end{figure}

\subsection{Wave functions}
Some wave functions are illustrated for a few values of $\gamma $ in Fig.2. The continuous change is clearly visible
when $\gamma $ is switched on. For real eigenvalues, in line with the underlying ${\cal PT}$-symmetry, the real part of 
the wave function must be symmetric and the imaginary part accordingly antisymmetric. This is the case when the factor $-i$ is
applied as discussed at the end of section 2. Note the increasing imaginary part of the wave functions 
associated with the largest and smallest eigenvalues for increasing $\gamma $ while it is the real part that is increasing
for the intermediate eigenvalue. Near to the EP3 the wave functions become essentially equal as expected. 
Since we deal with a non-hermitian Hamiltonian we must use the c-norm given by $\langle \tilde \Psi|\Psi \rangle$ (with
$\langle \tilde \Psi |$ being {\em not} the complex conjugate of $|\Psi \rangle $ but, in this case, rather its equal).
Also note that the norms, when taken separately for the real and imaginary part, become comparable for 
$\gamma \approx \gamma_{EP3}$; in fact their respective c-norms (being the difference of the respective separate norms) 
vanish at the EP3 as can be noticed by the increasing scale of the wave function.

\begin{figure}[h!]
\begin{center}
\vspace*{-3mm}
\includegraphics[width=0.4\columnwidth]{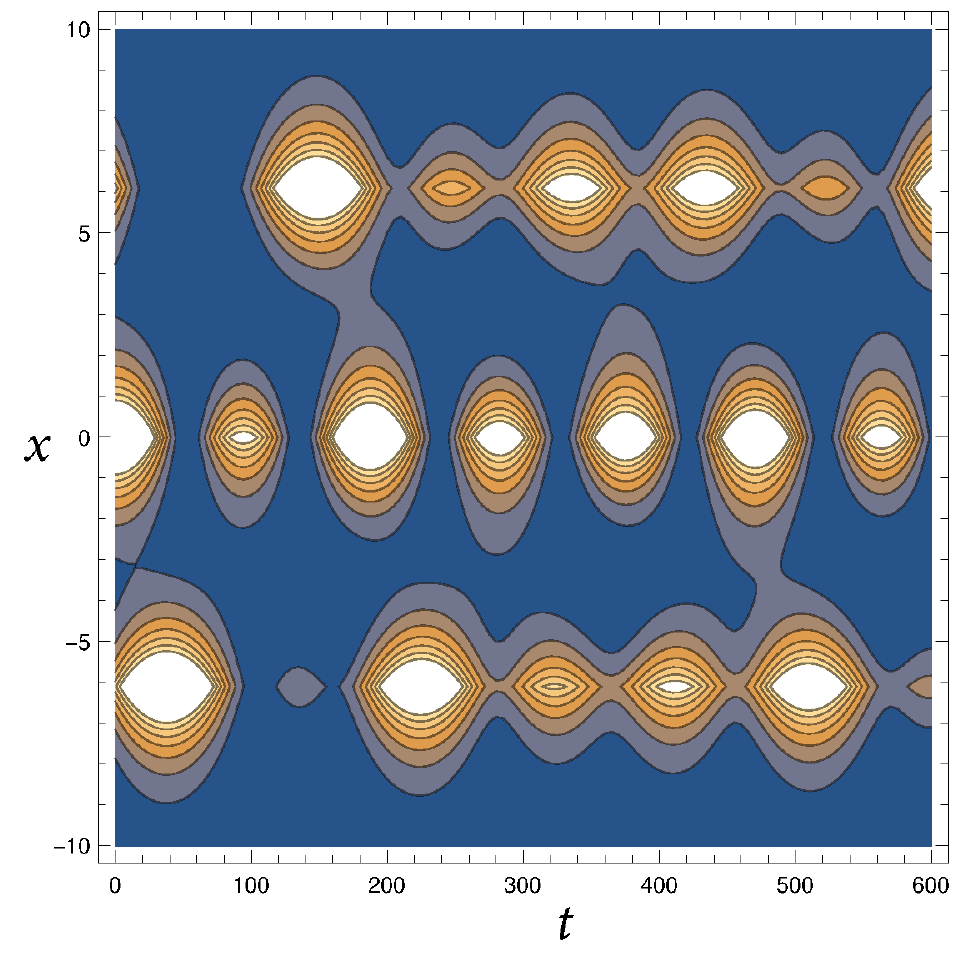}
\includegraphics[width=0.4\columnwidth]{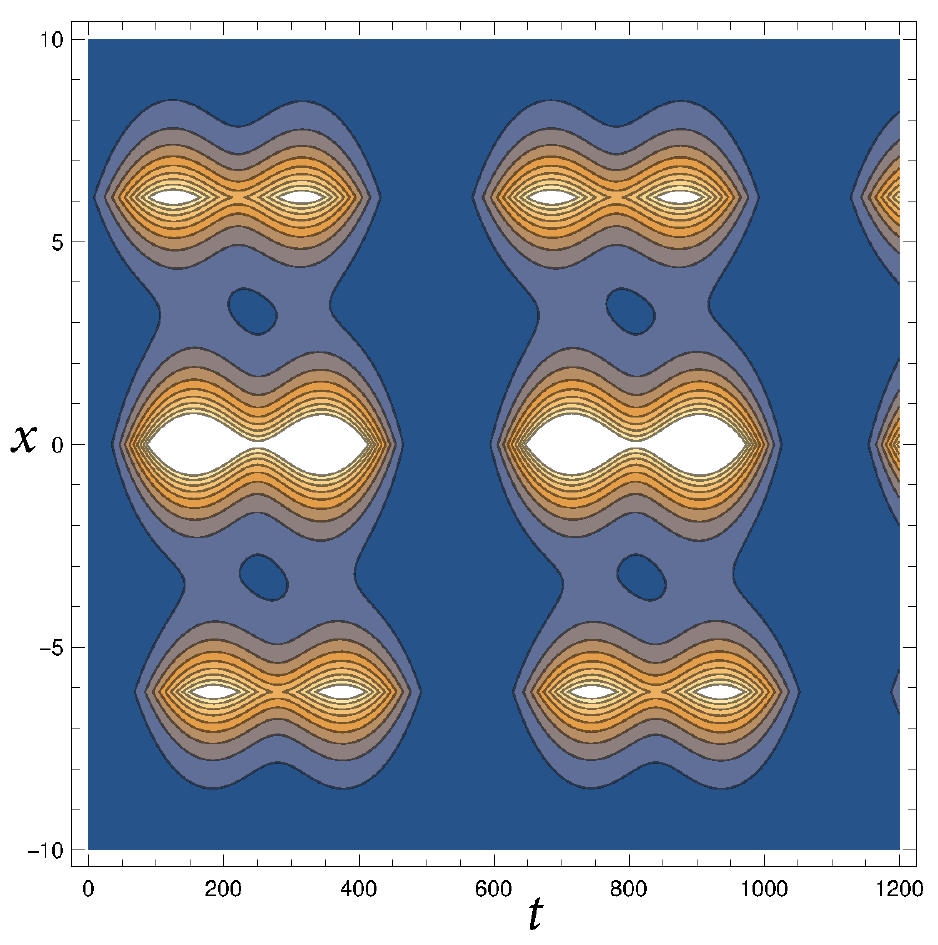}
\includegraphics[width=0.4\columnwidth]{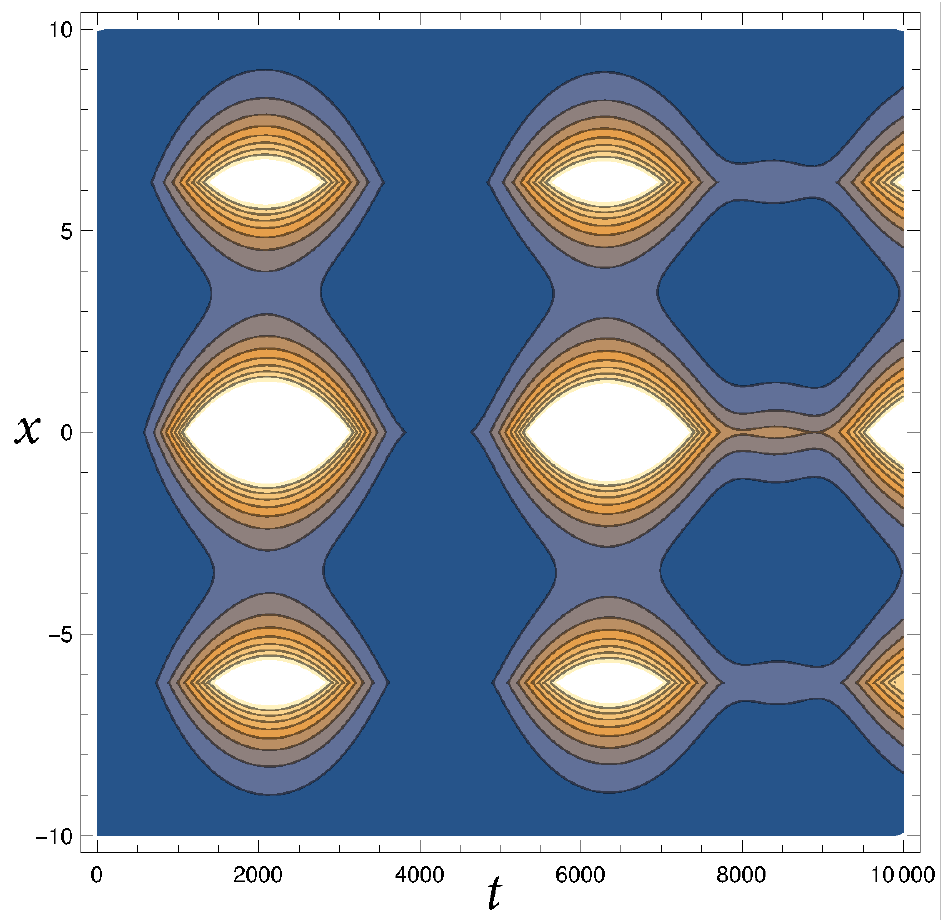}
\end{center}
\caption{Temporal evolution of a superposition of the three stationary solutions 
for increasing non-hermiticity parameter. Top left: $\gamma = 0, b = 6.1$,
height of contour lines from 0.6 to 0.1, top right:
$\gamma = 0.06, b = 6.1$, height of contour lines from 9 to 1,  
bottom: $\gamma = 0.65, b = 6.2075$, close to the EP3,
height of contour lines from 16000 to 2000.}
\label{fig_3}
\end{figure}

\subsection{Time evolution}
The time evolution of the wave functions is given by
\begin{equation}
|\Psi(t,x)\rangle=\sum _{i=1}^3 c_i\frac{|\Psi_i(x)\rangle}{\langle \tilde \Psi_i|\Psi _i\rangle }\exp (i\,k_i^2\,t)
\label{wavetime}
\end{equation}
with $c_i=\langle \tilde \Psi _i|\Psi(t=0)\rangle $ determining an initial condition.
Note that here the c-norm is of utmost importance. The modulus squared of (\ref{wavetime}) yields the intensity within
the respective potential wells as a function of $t$ and $x$. The illustration in Fig.3 displays the results for different
parameter sets.

Note the different time scales. Also note the different scales of the heights of the contours owing to the decreasing c-norm.
What becomes immediately obvious is the increase of the repeat time of the maxima with
increasing $\gamma $. Note that we cannot expect strict periodicity as the three eigenvalues are likely to be incommensurate.
What is observed here is the beat produced by the three eigenenergies. They are fairly distant for $\gamma =0$ yielding
a short beat time. In contrast, they
are very near to each other for $\gamma =0.065$ yielding a very long beat time. 
Except for $\gamma =0$ the pattern is virtually independent of
some initial conditions: the system has 'its own life', irrespective of the way it is triggered. This aspect is 
particularly pronounced in the vicinity of the singularity. We mention that very close to the EP3 the detailed pattern depends
somewhat on the path in the two parameter plane by which the EP3 is approached while the gross features prevail.

For different choices of $b$ and $\Gamma $ an EP3 can be found up to $\Gamma = 1.11$. For values of $\Gamma > 1.002$,
$\gamma_{EP3}$ becomes larger while $b_{EP3}$ becomes smaller. Values for $\Gamma = 1.11$ are:
$\gamma_{EP3}= 0.64, \;b_{EP3}= 2.40,\;k_{EP3}= 0.30$ (the exact values have more digits). 
However, the qualitative patterns remain similar in that
substantial changes happen only close to the EP3. For small values of $b$ (and $\gamma \ll \gamma_{EP3}$), 
the levels are rather distant due to the stronger
coupling. This is why a larger value of $\gamma $ is needed to force the three levels together. But again, the levels
depend weakly on the increasing $\gamma $, that is Fig.3 remains qualitatively unchanged.

\section{\label{concl}Summary and outlook}
Using a simple ${\cal PT}$-symmetric model for the interaction of three wave guides an exceptional point of third
order can be identified for a certain parameter range. Some of the features found are qualitatively reminiscent
of similar investigations for two interacting wave guides and an EP of second order. The time dependent pattern (being actually
the mode pattern along the extension of the wave guides) shows the characteristic pattern of more and more distant
intensity maxima when the spectral singularity is approached.

The new aspect of the present paper is the much increased sensitivity of parameter dependence in the approach of
the EP3. Moreover, in the close approach of the EP3 the three eigenvalues cannot be real if only one real parameter
is tentatively used. In fact, if an experiment would be attempted with only one parameter two of the three closely 
lying eigenvalues would infallibly form an EP2 and then disappear into the complex plane. In other words, at least
two parameters must be tuned carefully to force three {\em real} eigenvalues into an EP3 and thus visualize 
the expected pattern. While this constitutes
a great challenge for experimentation we feel that the present paper could give some guidance.

Along this line, an experimental verification of an EP of even higher order would be accordingly more demanding.

\appendix
\section{Explicit coefficients}
\label{coeff}
Setting $r=1$ we read off from (\ref{lineq})
\begin{eqnarray}
\varrho_{1}^{(i)} = -\frac{(1 + i \gamma )\exp(-2 b k_i)+  1 + i \gamma - 2 k_i}
{(1 + i \gamma ) \exp(-2 b k_i) -1 - i \gamma + 2 k_i} \\
\varrho_{2}^{(i)} = +\frac{(1 - i \gamma )\exp(-2 b k_i)+  1 - i \gamma - 2 k_i}
{(1 - i \gamma ) \exp(-2 b k_i)-1 + i \gamma + 2 k_i},
\end {eqnarray}
where a zero $k_i,\,i=1,2,3$ of the determinant of ${\cal M}$ has to be inserted to satisfy the third equation of (\ref{lineq}). 
Note that both denominators vanish at the intermediate (the second) value $k_2$ 
for $\gamma \to 0$, i.e.~when  $b=-{\rm Log}(1-2k_2)/(2k_2)$.
Inserting this expression into $\varrho_{1}^{(2)}$ and $\varrho_{2}^{(2)}$ the expansions in powers of $\gamma $ 
as given in the main text follow.

\section{Relation to realistic wave guides}
\label{waveguides}
Let us assume a wave guide of width $a$ centred around the origin $x = 0$ and placed on a 
substrate with background refractive index $n_0$. Across the wave guide the refractive 
index is altered to $n(x) = n_0 + \Delta \tilde{n}$, where $\Delta \tilde{n}$ is
complex, constant, and $| \Delta \tilde{n} |\ll n_0$.
The eigenvalue equation of the amplitude of the electric field vector of an
electromagnetic wave propagating along the wave guide in the $z$ direction, 
${\tilde E}_y(x,z,t) = E_y(x) {\rm e}^{i (\beta z - \omega _0 t) }$, with $\beta$ the propagation constant,
reads ({\it cf.} \cite{klaiman08a}) 
\begin{equation}
\left( \frac{{\rm d}^2}{{\rm d}x^2} + k_0^2 n^2(x) \right ) E_y(x)= \beta^2 E_y(x) \; ,
\label{eigenvalue_eq_E}
\end{equation}
where $k_0 = 2 \pi/\lambda_0 = \omega _0/c $, with $\lambda_0$ the vacuum wavelength.
To first order in $\Delta {\tilde n}$, $n^2(x) = n_0^2 +2 n_0 \Delta {\tilde n}$, and 
(\ref{eigenvalue_eq_E}) becomes
 \begin{equation}
\left( \frac{{\rm d}^2}{{\rm d}x^2} + 2 n_0 \Delta {\tilde n} k_0^2  \right ) E_y(x)= 
(\beta^2  - n_0^2 k_0^2) E_y(x) \; .
\label{eigenvalue_eq_E_2}
\end{equation}
If we  replace the effect of the altered refractive index across the wave guide
by a delta function $V_0 \delta(x)$ we have to require that the integral across the wave guide obeys
$2 n_0 \Delta {\tilde n} k_0^2 \; a = V_0$. Instead of (\ref{eigenvalue_eq_E_2}) we then have
\begin{equation}
\left( \frac{{\rm d}^2}{{\rm d}x^2} + 2 n_0 \Delta {\tilde n} k_0^2 \; a \; \delta(x) \right ) E_y(x)= 
(\beta^2  - n_0^2 k_0^2) E_y(x) \; .
\label{eigenvalue_eq_E_3}
\end{equation}

We decompose $\Delta {\tilde n}$ into its real and imaginary parts, $\Delta {\tilde n} =
\Delta n + i \Delta n^\prime$. The characteristic length scale set by $\Delta n$  is
\begin{equation}
\ell   = (2 n_0 \Delta {n} k_0^2)^{-1/2}. 
\end{equation}
Defining a further length scale $ L = \ell^2/a$ and
multiplying (\ref{eigenvalue_eq_E_3}) by $-L^2$, by noting that
$L\; \delta(x) = \delta(\tilde x)$,
with the dimensionless coordinate
${\tilde x} = x/L$, we finally obtain
\begin{equation}
\left( - \frac{{\rm d}^2}{{\rm d}{\tilde x}^2} - (1 + i \Delta n^\prime / \Delta n) \;\delta({\tilde x}) \right ) E_y({\tilde x})= 
- L^2(\beta^2  - n_0^2 k_0^2) E_y({\tilde x}) \; .
\label{eigenvalue_eq_E_4}
\end{equation}
Comparing (\ref{eigenvalue_eq_E_4}) with the triple delta-function Schr\"odinger equation 
(\ref{SE}), we  
identify $\gamma = \Delta n^\prime / \Delta n $, where the eigenvalues of 
(\ref{eigenvalue_eq_E_4})
are related to the eigenvalues $k^2$ of 
(\ref{SE}) by $k^2= L^2(\beta^2  - n_0^2 k_0^2)$,
and the dimensionless coordinate $x$ in (\ref{SE}) is equal to $\tilde x$.

To give an example, for the values used in \cite{klaiman08a}, $n_0 = 3.3, \Delta n = 10^{-3}$,
$ \lambda = 1.55 ~ \mu {\rm m}$ and $a = 5 ~ \mu {\rm m}$, one obtains $\ell = 3.036 ~ \mu {\rm m}$
and $ L = 1.843 ~ \mu {\rm m}$, both on the order of
the wavelength of the injected microwave.

\section*{Acknowledgement}
WDH and GW gratefully acknowledge the support from the National Institute for Theoretical Physics (NITheP), 
Western Cape, South Africa. GW expresses his gratitude to the Department of Physics of the University of
Stellenbosch where this work was carried out.

\section*{References}

\end{document}